# Pattern Formation without Favored Local Interactions

## (Abbreviated Title: "Ensembled Cellular Automata")

Alexander D. Wissner-Gross

Department of Physics, Harvard University, Cambridge, Massachusetts 02138 USA

Phone: +1-516-729-9726, Email: alexwg@physics.harvard.edu

**ABSTRACT**

Individual cellular automata rules are attractive models for a range of biological and physical self-assembling systems. While coexpression and coevolution are common in such systems, ensembles of cellular automata rules remain poorly understood. Here we report the first known analysis of the equally weighted ensemble of all elementary cellular automata (ECA) rules. Ensemble dynamics reveal persistent, localized, non-interacting patterns, rather than homogenization. The patterns are strongly correlated by velocity and have a quasi-linear dependence on initial conditions. Dispersion from a single initial site generates peaks traveling at low-denominator fractional velocities, some of which are not discernible in individual rules, suggesting collective excitation. Further analysis of the time-evolved rule space shows the 256 ECA rules can be represented by only approximately 111 principal components. These results suggest the rather surprising conclusion that rich self-assembly is possible without favoring particular local interactions.

# 1 INTRODUCTION

There has been much recent interest in programmable self-assembly of biological and material components [1–3]. Two very general features of self-assembly are particularly noteworthy for such applications. First, for a collection of $n$ components, each of which can take one of $k$ states, the total number of possible (not necessarily local) instantaneous interactions is $k^{(k^n)}$, since the future state of each component can be determined by any subset of the collection of current component states. In contrast, the total number of configurations of the collection is only $k^n$, and so it may be concluded that interaction spaces are generically larger than configuration spaces. Second, self-assembly often necessitates multiple cooperating or competing interactions as seen, for example, in genetic coexpression [4] and hydrophobic-hydrophilic protein folding [5]. It would therefore be interesting if an entire interaction space could be systematically surveyed to eventually enable rational tuning of multiple interactions for controlled self-assembly.

Cellular automata are attractive models for such self-assembly processes [6]. In particular, the 256 rules of elementary cellular automata (ECA) [7] are a model class of local interactions whose approximations and statistical behavior have been studied in detail [8-9]. It has been found that single iterations under the elementary cellular automata rule set are approximately linearized by a surprisingly small number of principal components [9]. However, single iterations do not capture the rich behaviors of cellular automata that require feedback between sites [7]. In the present work, in order to better understand the co-expressive behaviors of cellular automata rule spaces, we study for the first time the co-evolution of common initial configurations under ensembles of entire rule classes. In contrast to single-iteration studies [9], the ensembles considered in this paper consist of multiple applications of individual rules, which introduce nonlinearity that has not, to the best of our knowledge, been studied previously.

Rule ensembles may not have received significant attention previously because they typically include non-quiescent rules [10] that invert null initial states. (In this work, periodic boundaries are imposed, in part to avoid boundary artifacts from such non-quiescent rules.) While ensembles can certainly be approximated by typically large samples of stochastic cellular automata, stochastic ensembles of small finite state automata have received the most scrutiny [11]. Finally, an ensemble of ECA can be embedded in a $2^{256}$-color automaton, so perceived complexity may also have discouraged previous studies.

## 2 DEFINITION

In this work, we will focus primarily on the mean evolution of entire classes of transition rules, and in particular the ECA class. We define the equal weighting of ECA rules at position $x$ and iteration $n$ as

$$e_{n,x} \equiv \frac{1}{256} \sum_{k=0}^{255} s_n^k(x),$$

where $s_n^k(x)$ is the value at position $x$ and iteration $n$ of an elementary cellular automaton obeying transition rule $k$ (following Wolfram's notation [7]) that acts on a periodic configuration of size $L$. For simplicity, our early discussion assumes an initial configuration of $1\overbrace{000\cdots000}^{L-1}$. However, in later discussion, other initial configurations and transition rule classes are considered, while maintaining an equal rule weighting.

## 3 RESULTS AND DISCUSSION

Consider the evolution of a single site, averaged with equal weighting over all elementary rules, as shown in Figure 1(a). The evolution is eventually periodic with period 2, because that is the least common multiple of cycle sizes over the 4 possible 2-state automata rules. The same periodicity is reflected in the alternating behavior of backgrounds under non-quiescent rules [7]. Because the rule set is symmetric in site replacement values, the ensemble takes the value 1/2 at iteration 1. For subsequent even iterations, the ensemble takes the value 3/4 because 3 of the 4 possible 2-state automata are stationary after a single iteration and the remaining automaton is a 2-cycle that returns to its initial state after two iterations. Uniform-valued initial configurations must also have this ensemble oscillatory behavior by translational symmetry.

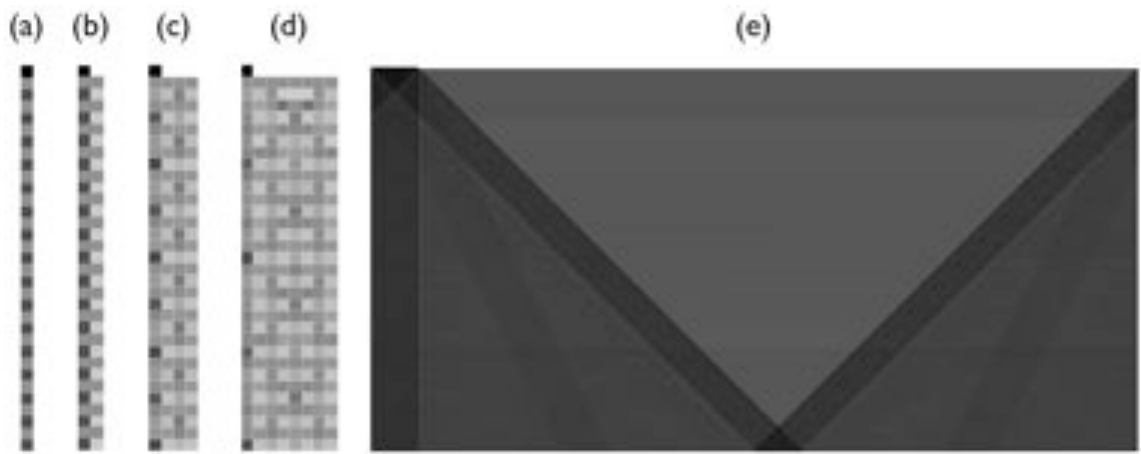

FIGURE 1

Visualization of evolution of configurations of different sizes. (a-d) Evolutions for time $T=32$ of initial configurations with a single nonzero initial site and sizes (a) 1, (b) 2, (c) 4, and (d) 8. (e) Evolution of 32 adjacent nonzero sites in a configuration of size $L=512$ over time $T=256$. Contrast is enhanced to reveal the presence of fronts with speeds 0, $\pm c/2$, $\pm c$.

The downwards direction corresponds to successive iterations.

More interesting behavior is observed in the evolution of initial configurations of larger sizes consisting of a single nonzero site. For configuration sizes $L$=2,4,8, as shown in Figure 1(b-d), propagation fronts traveling at the maximum allowed velocities $\pm c$ are discernible against a background 2-cycle. Surprisingly, the propagation fronts reach finite asymptotic amplitudes, and do not weaken through intersection, despite the fact that few elementary rules are linear [12]. Moreover, in the evolution of large configurations, faint fronts traveling at velocities $\pm c/2$ are also visible, as seen in Figure 1(e). Such narrow peaks do not appear to be present in any of the individual cellular automata evolutions from individual sites [12]. These well-defined structures beg the question of what other collective velocities, if any, are represented in the dispersion of elementary cellular automata ensembles, which we will revisit shortly.

But let us first examine the asymptotic behavior of these ensembles, since they appear not to homogenize, as might be naively expected for the average of "uncorrelated" discrete evolutions. The fraction or density of sites at iteration $n$ with value 1, $\lambda(n)$, is a useful statistical measure of a cellular automaton's equilibration [7]. Here, we will define $\lambda(n)$ as additionally averaging over an entire rule set. As visible in Figure 1(b-d), for a range of configuration sizes, $\lambda(n)$ follows approximately a 2-periodic orbit. The even-iteration value in the orbit appears to approach 1/3 with increased configuration size, while the odd-iteration value approaches ~0.531, as can be seen in the large-configuration limit of Figure 2(a-b). After $L/c$ iterations, the even- and odd-iteration density attractors both remain within 2% of their respective asymptotic averages. An odd-iteration attractor close to 1/2 is expected because that is the density in early odd iterations in the limit of large, nearly empty configurations. However, the origin of these particular limit values is still anomalous. Equilibration occurs partially after *L/(2c)* iterations and almost completely after approximately *L/c* iterations, consistent with the times required for fronts with speed *c* to first cross each other and then traverse the entire configuration. Additionally, Figure 2(a-b) confirms that the fronts are wave-like and spatially localized, since the densities do not converge to 1.

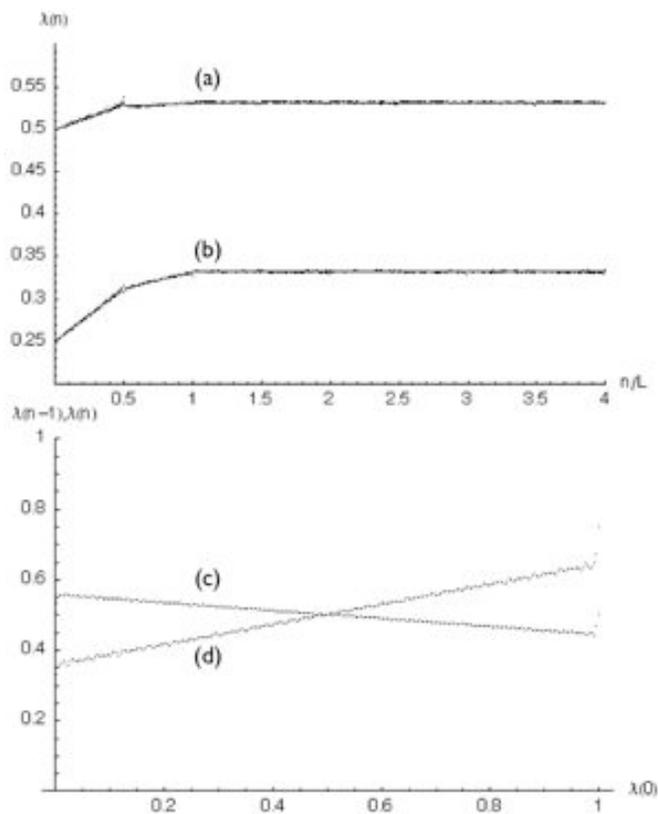

FIGURE 2

Equilibration of ensemble patterns. (a-b) Evolution of density $\lambda(n)$ for (a) odd and (b) even *n* (*L*=512, *T*=2048). (c-d) Equilibrium densities at (c) odd and (d) even iterations from initial configurations with varying density $\lambda(0)$ (*L*=256, *T*=256.)

By examining the evolution from initial configurations of varying densities, it is also possible to separate out the asymptotic contribution of the propagation fronts from the background, as shown in Figure 2(c-d). The asymptotically linear interference of the propagation fronts is reflected by the quasi-linearity of both curves. This quasi-linear density dependence is somewhat surprising given the sensitive, non-linear dependence on initial conditions of many individual rules. It should be noted, however, that initial densities $\lambda(0)$ near 0.0 or 1.0 lead to asymptotic densities $\lambda(T),\lambda(T-1)$ that diverge from the linear trend. For example, while $\lambda(T) \to 1/3$ as $\lambda(0) \to 0^+$, $\lambda(T)=1/4$ for $\lambda(0)=0$.

Now that we have discuss the non-interaction of the fronts, we return to our analysis of the front velocities, motivated by the visibility of faint $\pm c/2$-velocity fronts in ensembles. In particular, let us examine the ensemble that evolves from a single site over time $T >> L/c$ (i.e., after the primary $\pm c$ fronts have crossed many times), according to the correlation measure,

$$f_{0;1}(v) = \frac{2}{T-1} \sum_{\substack{n \text{ even},\ 0 \leq n \leq T-2; \\ n \text{ odd},\ 1 \leq n \leq T-1}} \left[ \left( \lfloor vcn \rfloor + 1 - vcn \right) \cdot e_{n,\lfloor vcn \rfloor} + \left( vcn - \lfloor vcn \rfloor \right) \cdot e_{n,\lfloor vcn \rfloor+1} \right],$$

where $v$ is the fractional velocity and $e_{n,x}$ is the mean ensemble value at iteration $n$ and displacement $x$ from the single nonzero initial site. Note that linear interpolation was used in the above measure to avoid aliasing artifacts. The resulting velocity spectra over odd and even iterations are shown in Figure 3(a-b). Further interference modes with low-denominator fractional speeds become evident, the most prominent being $c/5$, $2c/5$, $3c/5$, and $4c/5$ on even and odd iterations and $c/3$ and $2c/3$ on odd iterations. We believe that the origin of such well-defined collective speeds in the mean ensemble over the entire ECA rule class is worthy of further investigation.

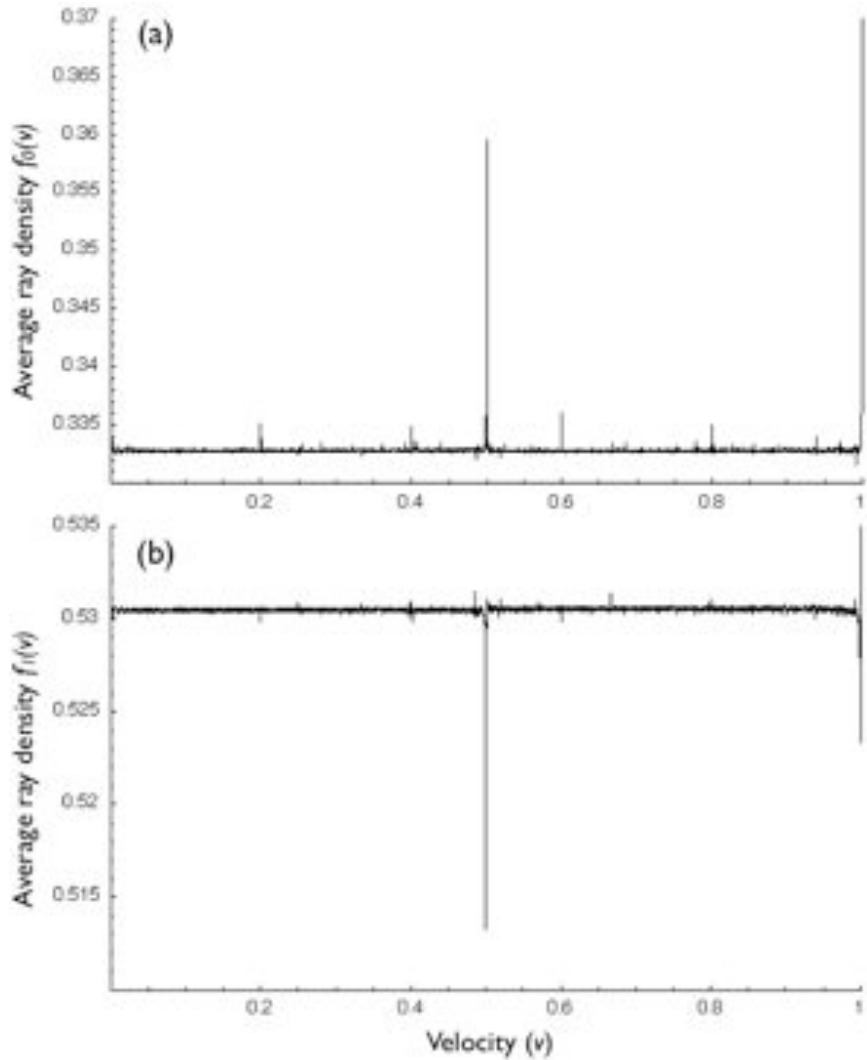

FIGURE 3

Dispersion spectra as measured by average density radiated outward from a single site at each velocity over (a) even and (b) odd iterations. ($L=256$, $T=32768$.)

There is a second natural way to decompose the ECA ensemble besides by velocity in the equally weighted sum over all rules—namely, by principal components. In our principal component analysis, the space-time evolutions of rules are treated as vectors, which undergo an orthogonal linear transformation to a new coordinate system with the property that the greatest variance of the vector set is parallel to the first coordinate, the second greatest variance of the vector set is parallel to the second coordinate, and so on [13]. Principal component analysis can therefore reveal reduced-dimensional representations of the ECA rules. In fact, the persistent non-interacting structures in the ensemble mean already suggested the presence of simple linearly independent components as a partial basis set. A full principal component set for the ECA over a finite time and configuration size is shown in Figure 4. In the principal component analysis of a configuration with $L=16$ over time $2L/c$, starting from a single nonzero site, 111 components were found, a surprisingly small count compared to the number, 256, of ECA rules. A sharp dropoff in eigenvalue is visible in the 5 most dominant principal components, which appear geometrically simple and feature the $\pm c$ fronts and the even/odd iteration alternating densities. The next 9 components featured additional structure propagating at speeds of $\pm c$, but still appear simple at large scales. The presence of at least some geometrically simple principal components in the ECA may prove useful for attempts at self-assembly based on linearly combining cellular automata rules.

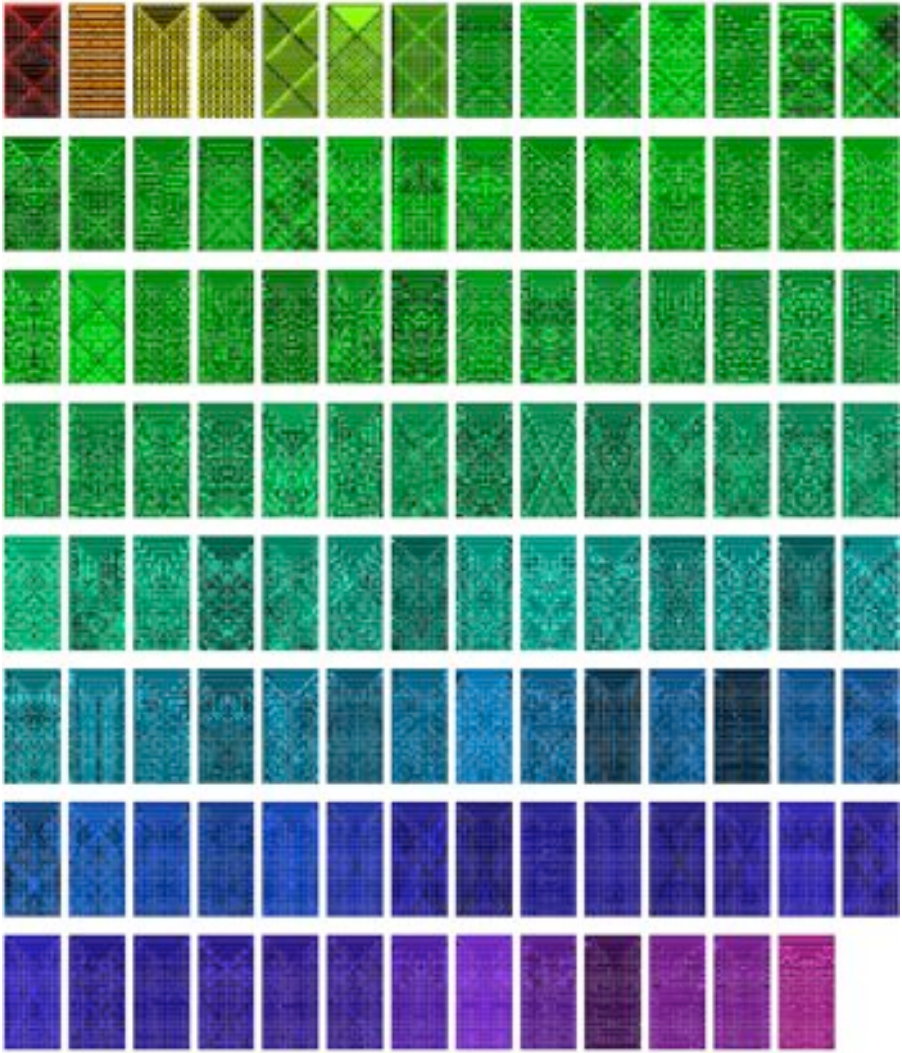

FIGURE 4

Principal component decomposition of the elementary cellular automata rules. The 111 principal components for the ensemble ($L$=16, $T$=32) are ordered by decreasing eigenvalue (first principal component is red), first from left to right, then top to bottom. The hue of each component represents the scaled logarithm of that component's eigenvalue, and the sharp dropoff of eigenvalues in the first 5 components is visualized as a comparatively rapid transition from red to green as compared to the finer spectrum of the remaining components.

We now conclude our study of linear ensembles of CA rule classes by demonstrating that the rich behaviors observed thus far are not limited to the elementary cellular automata, but are, in fact, quite generic and present in a variety of other rule classes as well. As shown in Figure 5, rich patterning beyond simple traveling fronts appears to be present whenever, for example, more than one nearest neighbor affects the outcome of the rules for 2 colors, but requires only one nearest neighbor for 3 colors. This requirement suggests that there is a minimum threshold of rule class complexity needed to produce nontrivial patterning in the ensemble mean that depends on both the number of colors and the neighborhood size. It should therefore not be surprising that similar patterning is observed in the 2-dimensional, 5-nearest-neighbor case as well, as shown in Figure 6.

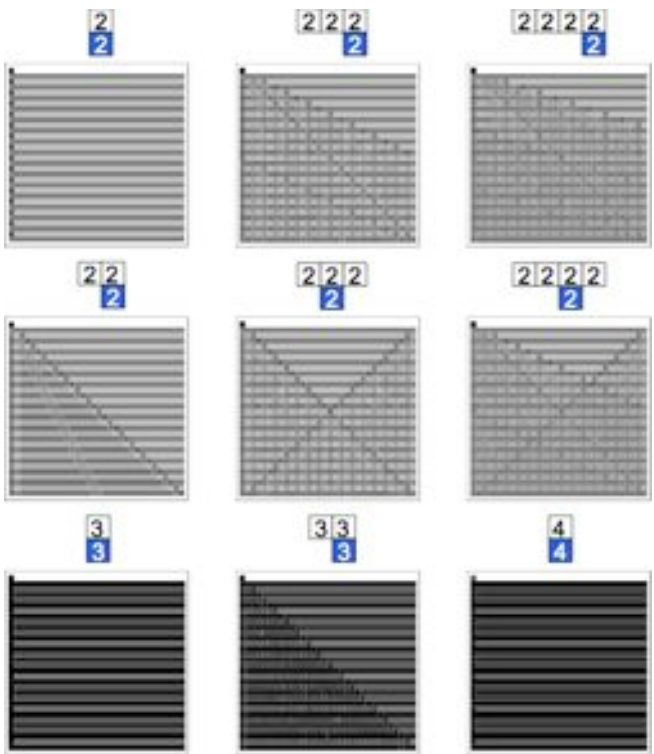

FIGURE 5

Ensembles means of other 1-D rule classes, with various numbers of site colors and neighbor dependencies. The number of colors per site and the neighbor dependence are indicated above each ensemble. ($L$=32, $T$=32.)

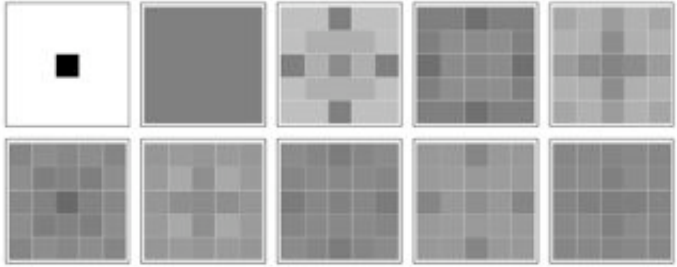

FIGURE 6

Ensemble mean of the first 10 iterations of all $2^{32}$ 2-dimensional, 5-nearest-neighbor rules.

## 4 CONCLUSION

We have shown, for the first time, that averages over spaces of iterated cellular automata rules produce rich interference structures rather than simple homogenization. A number of questions regarding the origin of the velocity and density of the structures remain open, and would be interesting to pursue in future work, since the mean ensemble we consider is such a natural encapsulation of the elementary cellular automata. Regardless, we have demonstrated with this simple system that rich patterning based on local interactions is possible without favoring any single interaction, which should be useful knowledge for future work in self-assembly.

## ACKNOWLEDGEMENT

The author thanks The Fannie and John Hertz Foundation for financial support, and E. Kaxiras, M. Stopa, G. Drescher, T. Sullivan, R. Nagpal, and S. Wolfram for helpful discussions.

## REFERENCES

[1] Nagpal, R. (2002). "Programmable Self-Assembly Using Biologically-Inspired Multiagent Control," *Proceedings of the 1st International Joint Conference on Autonomous Agents and Multi-Agent Systems (AAMAS'02),* 418-425.

[2] Liao, S., Seeman, N. C. (2004). "Translation of DNA Signals into Polymer Assembly Instructions," *Science*, *306*, 2072-2074.

[3] Flynn, C. E., Lee, S.-W., Peellee, B. R., Belcher, A. M. (2003). "Viruses as vehicles for growth, organization and assembly of materials," *Acta Mater.*, *51*, 5867-5880.

[4] Stuart, J. M., Segal, E., Koller, D., Kim, S. K. (2003). "A gene-coexpression network for global discovery of conserved genetic modules," *Science, 302*, 249-255.

[5] Berger, B., Leighton, T. (1998). "Protein folding in the hydrophobic-hydrophilic (HP) model is NP-complete," *Proceedings of the 2$^{nd}$ Annual International Conference on Computational Molecular Biology*, 30-39.

[6] Adamatzky, A. (1994). *Identification of Cellular Automata* (Taylor Francis, London).

[7] Wolfram, S. (1983). "Statistical mechanics of cellular automata," *Rev. Mod. Phys.*, *55*, 601-644.

[8] Moore, C. (1998). "Predicting nonlinear cellular automata quickly by decomposing them into linear ones," *Physica D*, *111*, 27-41.

[9] Deniau, L., Blanc-Talon, J. (1995). "PCA and cellular automata: a statistical approach for deterministic machines," *Complexity International*, *2*.

[10] Packard, N. H., Wolfram, S. (1984). "Two-dimensional cellular automata," *J. Stat. Phys.*, *38*, 901-946.

[11] Guinot, V. (2002). "Modelling using stochastic, finite state cellular automata: rule inference from continuum models," *Appl. Math. Modelling*, *26*, 701-714.

[12] Wolfram, S. (2002). *A New Kind of Science* (Wolfram Media, Champaign, IL).

[13] Jolliffee, I. T. (2002). *Principal Component Analysis* (Springer-Verlag, New York).